\documentclass[12pt,preprint]{aastex}

\shorttitle{Background Fluctuations at 4 $\mu$m}
\shortauthors{Xu et al.}

\begin{document}

\title{Measurement of Sky Surface Brightness Fluctuations at \[\lambda = 4 \mu m \]}

\author{Jing Xu}
\affil{California Institute of Technology, Mail Stop 59-33, Pasadena, CA 91125, USA}
\email{jingx@physics.ucsb.edu}

\author{James J. Bock}
\affil{Jet Propulsion Laboratory, Mail Stop 169-327, 4800 Oak Grove Drive, Pasadena, CA 91109, USA}
\email{jjb@astro.caltech.edu}

\author{Ken M. Ganga}
\affil{Infrared Processing and Analysis Center, Mail Stop 100-22, California Institute of Technology, Pasadena, CA 91125, USA}
\email{kmg@ipac.caltech.edu}

\author{Varoujan Gorjian}
\affil{Jet Propulsion Laboratory, Mail Stop 169-327, 4800 Oak Grove Drive, Pasadena, CA 91109, USA}
\email{vg@jpl.nasa.gov}

\author{Kazunori Uemizu}
\affil{Institute of Space and Astronautical Science, 3-1-1, Yoshinodai, Sagamihara-shi, Kanagawa 229-8510, Japan}
\email{uemizu@ir.isas.ac.jp}

\author{Mitsunobu Kawada}
\affil{Nagoya University, Chikusa-ku, Nagoya 464-8602, Japan}
\email{kawada@u.phys.nagoya-u.ac.jp}

\author{Andrew E. Lange}
\affil{California Institute of Technology, Mail Stop 59-33, Pasadena, CA 91125, USA}
\email{ael@astro.caltech.edu}

\author{Toshio Matsumoto}
\affil{Institute of Space and Astronautical Science, 3-1-1, Yoshinodai, Sagamihara-shi, Kanagawa 229-8510, Japan}
\email{matsumo@ir.isas.ac.jp}

\and 

\author{Toyoki Watabe}
\affil{Nagoya University, Chikusa-ku, Nagoya 464-8602, Japan}
\email{watabe@u.phys.nagoya-u.ac.jp}

\begin{abstract}
We present a measurement of faint source confusion in deep, wide-field 4 $\mu$m images.  The \( 1.8^{\circ} \times 1.8^{\circ} \) images with $17''$ resolution are centered about the nearby edge-on spiral galaxies NGC 4565 and NGC 5907.  After removing statistical noise and gain fluctuations in the focal plane array, we measure spatial fluctuations in the sky brightness to be \( \delta \nu I_{\nu} = 2.74 \pm 0.14 \) $nW$ $m^{-2}$ $sr^{-1}$, approximately 1\% of the diffuse background level observed in a single pixel.  The brightness fluctuations are confirmed to be associated with the sky by subtracting sequential images of the same region.  An auto-correlation analysis shows the fluctuations are well described by unresolved point sources.  We see no evidence for surface brightness fluctuations on larger angular scales ( \( 2^{\prime} < \theta < 28^{\prime} \) ). The statistical distribution of brightness fluctuations allows us to estimate the density of sources below the detection limit.  We present a Monte-Carlo analysis of the undetected sources, which appear to be dominated by galaxies.  Galaxies producing detected sky fluctuations contribute \( \nu I_{\nu} (>S) = 1.04^{+0.86}_{-0.34}\) $nW$ $m^{-2}$ $sr^{-1}$ to the cosmic infrared background, evaluated at \( S = 4.0 \times 10^{-8} \) $nW$ $m^{-2}$.  From the fluctuation data we can determine the integrated source counts \( N(>S) = 1.79^{+0.26}_{-0.40} \times 10^{7} \) $sr^{-1}$, evaluated at \( S = 4.0 \times 10^{-8} \) $nW$ $m^{-2}$.  The observed fluctuations are consistent with reddened $K$-band galaxy number counts.  The number counts of extracted point sources with flux \( \nu F_{\nu} > 6.3 \times 10^{-7} \) $nW$ $m^{-2}$ are dominated by stars and agree well with the Galactic stellar model of Wright \& Reese (2000).  Removing the stellar contribution from DIRBE maps with zodiacal subtraction results in a residual brightness of 14.0 $\pm$ 2.6 (22.2 $\pm$ 5.9) $nW$ $m^{-2}$ $sr^{-1}$ at 3.5 (4.9) $\mu$m for the NGC 5907 field, and 24.0 $\pm$ 2.7 (36.8 $\pm$ 6.0) $nW$ $m^{-2}$ $sr^{-1}$ at 3.5 (4.9) $\mu$m for the NGC 4565 field.  The NGC 5907 residuals are consistent with tentative detections of the infrared background reported by Dwek \& Arendt (1998), Wright \& Reese (2000), and Gorjian, Wright \& Chary (2000).

\end{abstract}

\keywords{ cosmology: observations --- diffuse radiation --- galaxies: individual (NGC 4565, NGC 5907) --- galaxies: statistical --- infrared: galaxies }

\section{Introduction}

Measurements of the cosmic infrared background (CIRB) probe the integrated emission from distant galaxies \citep{hau01}.  The brightness and spectrum of the CIRB is sensitive to galaxy number density, luminosity and evolution, as well as cosmology \citep{par67, kau76, sta80, dwe98, gis00}.  However, a search for the CIRB must account for the local zodiacal and stellar foregrounds.  These have proven difficult to eliminate, even at \( \lambda = 3 - 5 \) $\mu$m where the local foreground from thermal emission and scattering by zodiacal dust is minimal.

DIRBE reported upper limits on the CIRB at near-infrared wavelengths \citep{hau98}, limited by foreground subtraction.  The total astrophysical emission observed by DIRBE in a \( 0.7^{\circ} \) beam contains a significant emission component from stars.  Dwek \& Arendt (1998) used ground-based $K$-band galaxy counts along with the DIRBE data to obtain a tentative detection of the CIRB.  Gorjian, Wright \& Chary (2000) independently claimed a tentative detection of the CIRB using ground-based $L$-band counts to account for stellar emission in the DIRBE beams.  Wright \& Reese (2000) reported a detection using a different zodiacal model than used by Hauser et al. (1998).  Matsumoto (2000) analyzed observations with the Near InfraRed Spectrometer (NIRS) on IRTS, with an \( 8' \times 8' \) field of view that is significantly smaller than that of DIRBE, and reported an isotropic near-infrared component peaking at $\sim$ 1.5 $\mu$m.

Brightness fluctuations also probe the CIRB.  The fluctuations contain information on the number counts and clustering properties of sources below the detection limit \citep{bon86, fab88, col92, kas00}.  Poissonian fluctuations (``confusion noise'') are predicted based on models of galaxy evolution, including stellar and cirrus components from our galaxy \citep{fra91}.  The brightness distribution depends on the index, $\alpha$, of integrated source counts \( N(> S) = N_{0} \, {(S / S_{0})}^{- \alpha} \), and can thus be used to probe the density of sources below the detection limit \citep{con74}.  The spatial distribution of fluctuations measures the clustering of sources \citep{kno01} on size scales of galaxy clustering.  For example, NIRS/IRTS observations show brightness fluctuations prominent at \( \Delta \theta \approx 100' \), appearing at the level of $\sim$ 5\% of the total diffuse sky brightness \citep{mat00}.  Confusion noise sets the limiting sensitivity for future space-borne infrared observations, such as for SIRTF \citep{fan98} and Astro-F \citep{ona00}.

We present integrated source counts and a statistical analysis of residual brightness fluctuations in wide-field images obtained with a rocket-borne near-infrared camera at \( \lambda = 4 \) $\mu$m.  These images provide a direct account of the point source contribution to the near-infrared sky brightness, and an indirect measure of the brightness contribution from sources below this detection limit.

\section{Observations}

The near-infrared telescope experiment (NITE) consists of a 16 $cm$ rocket-borne telescope cooled by supercritical helium and a 256 $\times$ 256 low-background InSb focal plane array.  The array is located at the focus of an $f/2.2$ Gregorian telescope providing a \( 1.2^{\circ} \times 1.2^{\circ} \) field of view with a plate scale of $17''$ per pixel.  A 3.5 $\mu$m long-wavelength pass infrared filter and the intrinsic long wavelength cutoff of the InSb focal plane array defines the 3.5 to 5 $\mu$m spectral band of the instrument.  The array, sampled continuously and non-destructively ``up the slope'' at 3.8 $Hz$ over a 10 $sec$ reset interval, gives near-background-limited noise performance and maximum flexibility in data reduction.  The telescope and cryogenic baffle tube were cooled by supercritical liquid helium to eliminate thermal emission from the instrument.  The focal plane array was operated at $\sim$ 30 $K$ for optimal noise performance.  Further details on the instrument may be found in Bock et al. (1998).

NITE was designed to probe diffuse near-infrared emissions about nearby edge-on spiral galaxies, arising from possible low-mass stars in the massive halo.  Analysis of data from two flights, on 1997 May 28 centered about NGC 4565, and on 1998 May 22 centered about NGC 5907, produced strict upper limits on the near-infrared halo brightness \citep{uem98, yos00}.  For purposes of assessing point source contributions to the near-infrared background, the two fields allow us to evaluate the relative contributions of point sources in the NGC 4565 (\( b = 86^{\circ} \), \( l = 231^{\circ} \); \( \beta = 27^{\circ} \), \( \lambda = 177^{\circ} \)) and the NGC 5907 (\( b = 51^{\circ} \), \( l = 92^{\circ} \); \( \beta = 68^{\circ} \), \( \lambda = 188^{\circ} \)) fields.

We observed each region by stepping the field of view $0.6^{\circ}$ in a square pattern, and integrating for 20 $sec$ in two 10 $sec$ exposures at each position.  This sequence produces a complete map consisting of 8 individual exposures.  The observation sequence repeats with a slight shift of $4'$ to sample the sky with different pixels.  The central \( 32' \times 36' \) region around the target edge-on galaxy is thus observed in each of the 16 exposures.  Bright calibration stars, observed at the beginning and end of the sequence, and bright stars in the field allow us to calibrate the responsivity and point spread function (PSF) of the instrument. Further details on the observations for each flight may be found in Uemizu et al. (1998) and Yost et al. (2000).

\section{Data Reduction}

We present a statistical analysis from the 1997 observation about NGC 4565.  The 1998 observation of NGC 5907, at lower galactic latitude (\( b = 51^{\circ} \)), suffers from a relatively larger contribution from stars.  Furthermore, the 1998 observation has only 8 useful exposures due to a large terrestrial signal that appeared in the latter portion of the flight.  We use the 1998 observations to assess the relative abundance of detected point sources at lower galactic latitude.  The 1997 data has a noisy detector readout that samples every fourth column on the array.  The cause of the excess noise, a loose wire-bond to the focal plane array, was subsequently repaired after the flight.  One edge of the array was vignetted by a misalignment with the field stop.  We carefully omit data from the noisy column and the vignetted region throughout the fluctuation analysis (Sections 3.2, 3.3, 3.4, and 4.2).

We adopt the same reduction procedure as in Uemizu et al. (1998), briefly summarized as follows.  We correct each of the 16 exposures for dark current by subtracting a master dark frame with a median dark current of 1.55 $e^{-}/s$, obtained by closing an internal cryogenic (5 $K$) shutter just before launch.  The flat field response is derived by first normalizing each individual exposure on the sky to its mean value.  The mean is determined by applying an iterative 3$\sigma$ clip to remove point sources.  A flat field frame for each quadrant is then obtained by evaluating the median value in each pixel over 12 exposures.  Thus 4 flat field frames are generated for each of the 4 pointing positions used to observe the central galaxy.  The flat field for each pixel in this reduction procedure is thus derived from observations of numerous patches of sky separated by $\sim$ $36'$ from the observation position.

We divide all of the exposures by their corresponding flat field frames, and subtract a small constant value from each exposure to account for a slowly varying terrestrial signal.  The time-varying signal, $\sim$ 17 $e^{-}/s$, is small compared with the total photocurrent of 160 $e^{-}/s$.  The exposures are aligned to bright stars in the field, and combined to form a \( 1.8^{\circ} \times 1.8^{\circ} \) mosaic.  We constructed a star mask from the mosaic, masking all pixels 3$\sigma$ above the average over the image. Bright point sources are each masked out to regions such that their extended emission is no more than 2\% above the total sky background based on the PSF.  NGC 4565 is similarly masked based on the results of Uemizu et al. (1998).  As the combined mosaic has a higher sensitivity to point sources than the individual exposures, we re-evaluate the flat field frames using the same procedure as with the mask obtained from the mosaic.  The new flat field frames are then combined to produce a final mosaic.  

\begin{figure}
\plotone{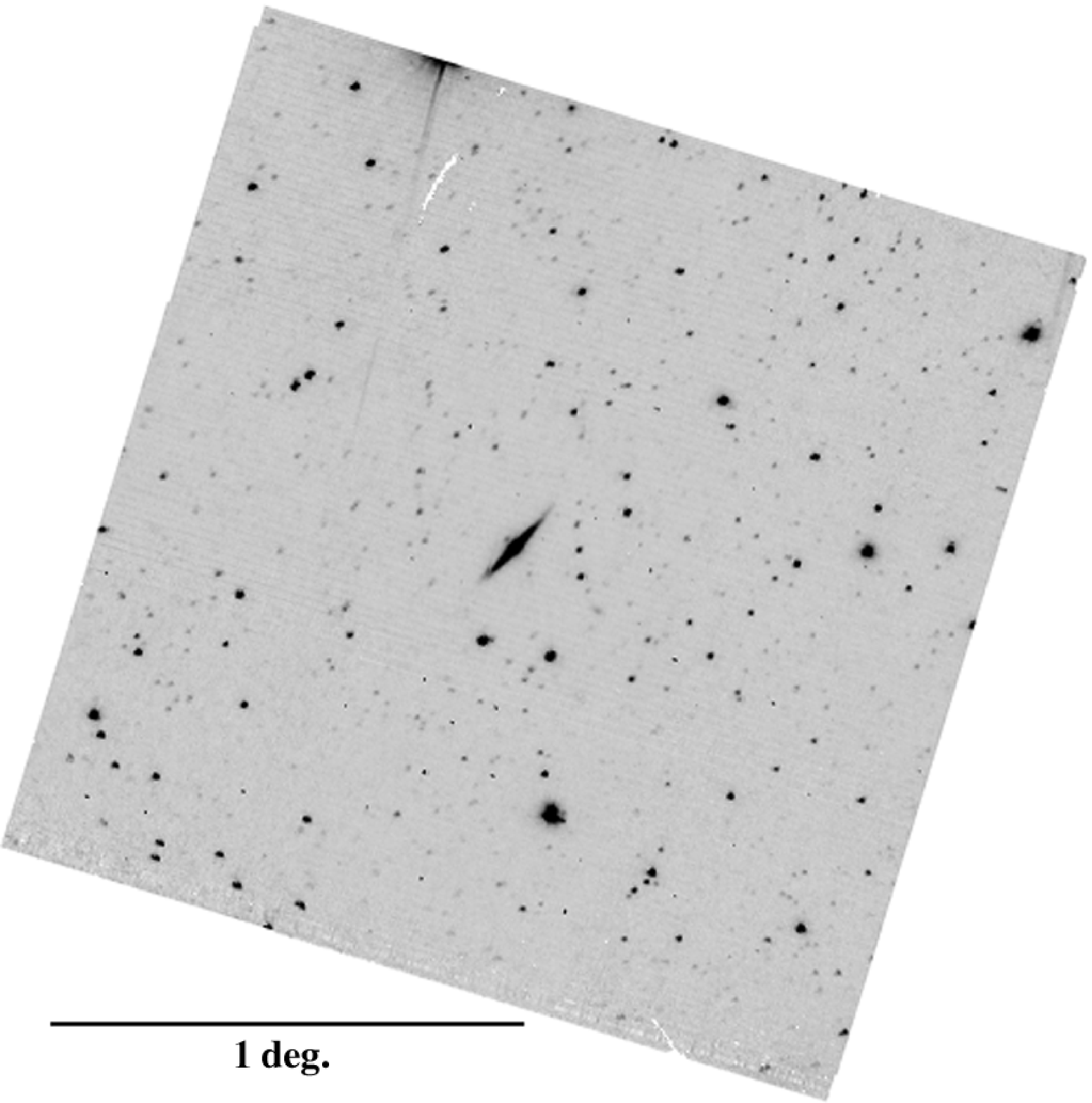}
\caption{ Gray scale image of the full \( 1.8^{\circ} \times 1.8^{\circ} \) field centered about the edge-on spiral galaxy NGC 4565.  The image is rotated so that equatorial north is up and east to the left.  Data from the noisy readout and the vignetted region are omitted in the fluctuation analysis (Sections 3.2, 3.3, 3.4, and 4.2).\label{fig:fov97}}
\end{figure}

The resulting gray-scale image is shown in Fig. \ref{fig:fov97}, including data from all four array readouts.  The regions on the corners of the image are obtained with only 2 pointings (4 exposures), while the central region of the image is well sampled with 8 pointings (16 exposures).  The image in Fig. \ref{fig:fov97} has prominent stripes due to data from the noisy column (for further evaluation of the statistical impact of the noisy columns see Fig 4).  The remaining 3 array readouts are separately binned and found to have very similar noise properties, with read noises of 52 $\pm$ 0.7 $e^{-}$, 53 $\pm$ 1 $e^{-}$, and 53 $\pm$ 0.7 $e^{-}$ for 10 $sec$ integration time.  We carefully omit data from the noisy column and the vignetted regions throughout the fluctuation analysis (Sections 3.2, 3.3, 3.4, and 4.2).  For the sake of simplicity, we confine the variation analysis in sections 3.2 and 3.3 to pixels with 100\% observing time.  To improve our statistical power, we include pixels with 75\% or more observing time for the analysis in sections 3.4 and 4.2.
 
\subsection{Extraction of Point Sources}

Identification of point sources in our field of view (Fig. \ref{fig:fov97}) is conducted with a PSF-fitting using DAOPHOT.  Bright calibration stars, observed at the beginning and end of the flight sequence, and bright stars in the field allow us to calibrate the responsivity and PSF of the instrument.  We iteratively extract sources by locating and masking the brightest sources in our field of view.  The identified sources are individually masked out to regions where the extended PSF brightness is $<$ 2\% of the mean sky brightness, \( \nu I_{\nu} = 260 \) $nW$ $m^{-2}$ $sr^{-1}$.  We use DAOPHOT to extract fluxes from identified sources.  For consistency check, we compare the DAOPHOT fluxes with aperture photometry.  In the case of very bright sources, we find some discrepencies, since a few are nearby extended galaxies.  We thus use aperture photometry for the bright and extended sources.  

The resulting integrated source counts, detected with $>$ 4$\sigma$ significance, are displayed in Fig. \ref{fig:num_count} for the 1997 and 1998 fields.  The number counts are not corrected for incompleteness at the faint end.  The large error bars at the bright end are due to the small number of sources.

\subsection{Evaluation of Sky Brightness Fluctuations}

We produce two independent sky images $S_{1}$ and $S_{2}$ from the first 8 and second 8 exposures respectively, to discriminate sources of noise from the instrument, sky, and variations in the gain matrix (``flat-field'' errors).  We estimate the displacement between $S_{1}$ and $S_{2}$ accurately by comparing the centroids of bright stars.  The resulting longitudinal and transverse displacements are $\leq$ 0.3 and 0.2 $pixel$ respectively.  As we shall see, subtracting the two images allows us to distinguish noise associated with the array, compared to fluctuations associated with a fixed pattern on the sky.

The measured noise in a single 10 $sec$ integration is 53 $e^{-}$, in good agreement with photon noise, \( (10 s \times 160 e^{-}/s)^{1/2} = 40 e^{-} \), and readout electronics noise, $\sim$ 30 $e^{-}$ as determined form dark frames before launch, combined in quadrature, \( ( (40 e^{-})^{2} + (30 e^{-})^{2})^{1/2} = 50 e^{-} \).  Instrument noise may also include various non-ideal noise sources at a low level, such as a fixed pattern noise on the array, low-level cosmic rays, incomplete subtraction of dark current, and time-varying signals.  Subtracting the $S_{1}$ and $S_{2}$ images, which are shifted in position relative to one another on the sky, therefore provides an accurate assessment of all sources of instrument noise which are necessarily correlated with array coordinates.

To evaluate sky fluctuations statistically, we confine our analysis to the 1997 field about NGC 4565.  $S_{1}$ consists of a mosaiced image from the first eight exposures, and $S_{2}$ consists of a mosaiced image from the second eight exposures.  We denote \( S_{sky} = (S_{1} + S_{2})_{sky} / 2 \), the average of $S_{1}$ and $S_{2}$ with respect to sky coordinates, as the combined sky image.  We mask $S_{1}$ and $S_{2}$ for 4$\sigma$ sources with a common mask derived from $S_{sky}$, taking advantage of the higher sensitivity to point sources in the combined image $S_{sky}$.  Both $S_{1}$ and $S_{2}$ are corrected for flat-field variations as described above with a common gain matrix.  We consider three sources of noise, instrument noise (defined to include all un-correlated noise between exposures), errors in the gain matrix, and fluctuations on the sky.

Because the gain matrix is determined from observations of the sky, it can introduce statistical correlations in the data.  We confine our analysis to the central \( 28' \times  28' \) region to eliminate statistical correlations and to minimize gain errors, ``flat-field noise'', as follows.  The majority of the pixels in the inner region of the combined image $S_{sky}$ are obtained with 8 independent pixels on the array.  The flat field for each of these pixels is obtained from 6 different patches of sky.  In constructing $S_{sky}$, each point in the inner region is compared with 48 patches of sky located $\sim 36'$ outside the inner region, 24 of which are independent.  In the limit that the combined image $S_{sky}$ is dominated by instrument noise, the variance of each pixel in the central region is increased by a factor of \( 8/48 = 1/6 \) due to flat-field variations.  In the limit that each exposure is dominated by sky fluctuations, the variance is increased by a factor of $1/24$.  As sky fluctuations and instrument noise are actually of similar amplitude in $S_{sky}$, the systematic variance of each pixel in the central region is increased by a factor of $1/10$.  The factor of $1/10$ is derived by noting that flat-field errors arise from the errors in the off beams, with instrument variance dominating over sky variance in the off beams by a factor of 4.  However, the variance in $S_{sky}$ is $\sim$ 2 times larger than instrument noise, so the relative flat-field variance is \( (5/4) (8/48) (1/2) \), or approximately $1/10$.  Thus variations arising from flat field errors should generally be small in comparison to sky fluctuations in the inner region, and do not contribute to statistical correlations for \( \Delta \theta < 28' \).

We assume the variance in \( S_{sky} = {(S_{1} + S_{2})}_{sky} / 2 \) arises from a combination of instrument noise ($v_{i}$), flat-field variations ($v_{ff}$), and sky fluctuations ($v_{s}$).  For the sake of simplicity, we further confine the following analysis to pixels with full observing time in the central \( 28' \times  28' \) region.  The measured variance in $S_{sky}$ is given by


\begin{equation}
v(S_{sky}) = v_{i} + v_{ff} + v_{s} = 15.04 \pm 0.69 \ nW^{2} \ m^{-4} \ sr^{-2} .
\label{eqn:varA}
\end{equation}

We note that sky fluctuations will have the same amplitude in $S_{1}$ and $S_{2}$ because $S_{1}$ and $S_{2}$ share a common flat field matrix.  However, the gain for each pixel is derived from an independent arrangement of ``off'' positions, so flat field variations are independent from pixel to pixel.  We assume that instrument noise is uncorrelated in the two images.  The variance in $S_{1}$ and $S_{2}$ is thus given as

\begin{equation}
v(S_{1}, S_{2}) = 2 v_{i} + 2 v_{ff} + v_{s} .
\end{equation}

We can separate these contributions by using the fact that $S_{1}$ and $S_{2}$ are displaced from one another by $4'$ on the sky.  By subtracting $S_{1}$ and $S_{2}$ with respect to sky coordinates, the variance in the differenced image \( D_{sky} = {(S_{1} - S_{2})}_{sky} / 2 \) becomes

\begin{equation}
v(D_{sky}) = v_{i} + v_{ff} = 7.55 \pm 0.35 \ nW^{2} \ m^{-4} \ sr^{-2} .
\label{eqn:varB}
\end{equation}

Comparing the variance of $S_{sky}$ and $D_{sky}$ thus gives a direct measure of the sky fluctuations, \( v_{s} = 7.49 \pm 0.77 \) $nW^{2}$ $m^{-4}$ $sr^{-2}$.

We evaluate the instrument noise contribution by subtracting image pairs viewing the sky 10 $sec$ apart, and determine the variance to be \( 185.90 \pm  1.5 \) $nW^{2}$ $m^{-4}$ $sr^{-2}$.  As 16 exposures comprise $D_{sky}$, the estimated contribution from instrument noise is 

\begin{equation}
v_{i} = \frac{1}{32}(185.90 \pm 1.5)\ nW^{2} \ m^{-4} \ sr^{-2} = 5.81 \pm 0.26 \ nW^{2} \ m^{-4} \ sr^{-2} .  
\label{eqn:varC}
\end{equation}

We then estimate the flat-field contribution to be \( v_{ff} = 1.74 \pm 0.44 \) $nW^{2}$ $m^{-4}$ $sr^{-2}$.  This result agrees within experimental error with our expectation that the variance from flat-field errors is \( v_{ff} = (v_{i} + v_{s})/10 = 1.33 \), as described above.  Furthermore, we can see that the instrument noise averages down with the number of exposures as expected.  If the array had a fixed pattern noise that did not average down, it would be accounted as flat-field noise.  However since the flat-field noise is in agreement with our expectation, such sources of excess noise must be small.  The variance in each pixel of the combined image $S_{sky}$ is thus marginally dominated by sky fluctuations, as shown in Fig. \ref{fig:histogram}.

We may test the reliability of our estimation by differencing the same two masked sky images with respect to array coordinates.  The resulting difference image \( D_{array} = {(S_{1} - S_{2})}_{array}/2 \) is free of flat-field variation, since $S_{1}$ and $S_{2}$ share a common gain matrix.  Its variance becomes

\begin{equation}
v(D_{array}) = v_{i} + v_{s}/2 = 10.50 \pm 0.55 \ nW^{2} \ m^{-4} \ sr^{-2} ,
\label{eqn:varCheck}
\end{equation}

\noindent
in agreement with our expected variance of 9.55 $\pm$ 0.60 $nW^{2}$ $m^{-4}$ $sr^{-2}$, obtained from eq. (\ref{eqn:varA}), (\ref{eqn:varB}), and (\ref{eqn:varC}) above.  This consistency check shows that flat-field errors, or a non-ideal fixed pattern noise associated with the array, are both small and accurately estimated.  It also shows that any effect from possible sub-pixel displacement between $S_{1}$ and $S_{2}$ is small comparing to the statistical uncertainty in our variance estimation, since eq. (\ref{eqn:varCheck}) does not require alignment on the sky between $S_{1}$ and $S_{2}$.

The detection of sky fluctuations can be shown graphically by evaluating the variance in \( D = (S_{1} - S_{2})/2 \) as a function of the positional shift between $S_{1}$ and $S_{2}$, as shown in Fig. \ref{fig:difplot}.  Pixels with 75\% or more observing time were used in Fig. \ref{fig:difplot} to get more complete coverage, although this slightly increases $v_{i}$ above the value obtained in eq. (\ref{eqn:varC}).  The variance is minimized when the two images are differenced according to sky coordinates, and does not appreciably decrease when differenced according to array coordinates (correspondingly marked by the ``+'' in the $\Delta$x -$\Delta$y plane, Fig. \ref{fig:difplot}).  The width of the minimum in the variance is consistent with the PSF of the instrument.  Therefore the observed sky fluctuations are unlikely to arise from improperly masking the extended emission of NGC 4565 or scattered light from bright stars.  As a check, we changed the masking level and found no significant change in the above results.  Also note that the striped pattern in the variance evaluation shown in Fig. \ref{fig:difplot}, an artifact arising from omitting data obtained with noisy readout columns, is small ($\sim$ 5\% in amplitude).

\subsection{Auto-Correlation Analysis}

We estimate the spatial structure of the sky fluctuations using an auto-correlation analysis using the central pixels with full observing time.  The two-point auto-correlation function \citep{gan93},

\begin{equation}
C(\theta) = <\delta \nu I_{\nu} ( x ) \cdot \delta \nu I_{\nu}(x + \theta)> ,
\end{equation}

\noindent
where \( \delta \nu I_{\nu} ( x) = (\nu I_{\nu} (x) - <\nu I_{\nu}>) \), can be used to probe fluctuations on angular scales from $17''$ to $28'$.  We evaluate the correlation function on the summed ($S_{sky}$) and differenced ($D_{sky}$) images, described in section 3.1.  A constant offset is removed from the summed image before auto-correlation analysis, obtaining an ensemble average of zero in $S_{sky}$.  Sky, instrument, and flat-field variance contribute to the variance of $S_{sky}$, whereas $D_{sky}$ contains no sky variance, only instrument and flat-field variance.  Since we only analyze the central \(28' \times 28'\) region, we can  remove any correlated structure as a result of flat-fielding.

The resulting correlation functions are shown in Fig. \ref{fig:autocorrelation}.  The correlation function at zero-lag, C(0), gives the variance resulting from instrument, sky fluctuation, and flat-field contributions as described in section 3.2.  The gradual rise in $C_{S_{sky}} (\theta)$ near $\theta = 0'$ is consistent with a combination of instrument noise and sky fluctuations arising from unresolved sources, spatially varying as the PSF of the instrument.  $C_{D_{sky}}(\theta)$ shows no gradual rise, as expected for variance arising from instrument noise.  We see no evidence for sky fluctuations at larger angular scales, \( 2' < \theta < 28' \).  We constrain \( \mid C_{S_{sky}}(\theta) \mid < 0.5 \) $nW^{2}$ $m^{-4}$ $sr^{-2}$ (2$\sigma$), averaged from \( 10' < \theta < 28' \).  Our non-detection of correlated structure cannot exclude the detection by Matsumoto (2000) on larger angular scales (\( \theta \sim 100' \)) and shorter wavelengths (1.4 $-$ 2.1 $\mu$m), where RMS fluctuations are significantly brighter.

We have shown the presence of statistical fluctuations that are in excess of the instrument and flat-field noise sources.  These fluctuations correlate with sky coordinates, not array coordinates.  Furthermore the correlation analysis shows these fluctuations are spatially described by point sources.  No known non-ideal noise source on the detector array can mimic these properties.  The correlation analysis also shows that sky fluctuations cannot arise from structure in the Zodiacal foreground.

\subsection{Histogram}

Further information on the nature of the sky fluctuations may be evaluated from the statistical ensemble of pixels in the central region of the image.  The brightness distributions of the 5000 central region pixels (with 75\% and more observing time) are shown in Fig. \ref{fig:histogram}, after masking the galaxy, and detected point sources.  The brightness distribution of summed image \( {(S_{1} + S_{2})}_{sky}/2 \), $H_{S_{sky}}$, is noticeably skewed towards positive sources, and significantly wider than that of the differenced image \( {(S_{1} - S_{2})}_{sky}/2 \), $H_{D_{sky}}$.  The differenced histogram $H_{D_{sky}}$ is well-described by a Gaussian, which eliminates noise sources on the array that could produce a non-Gaussian ``tail'' such as cosmic rays.

\section{Discussion}

\subsection{Source Counts}

\begin{figure}
\plottwo{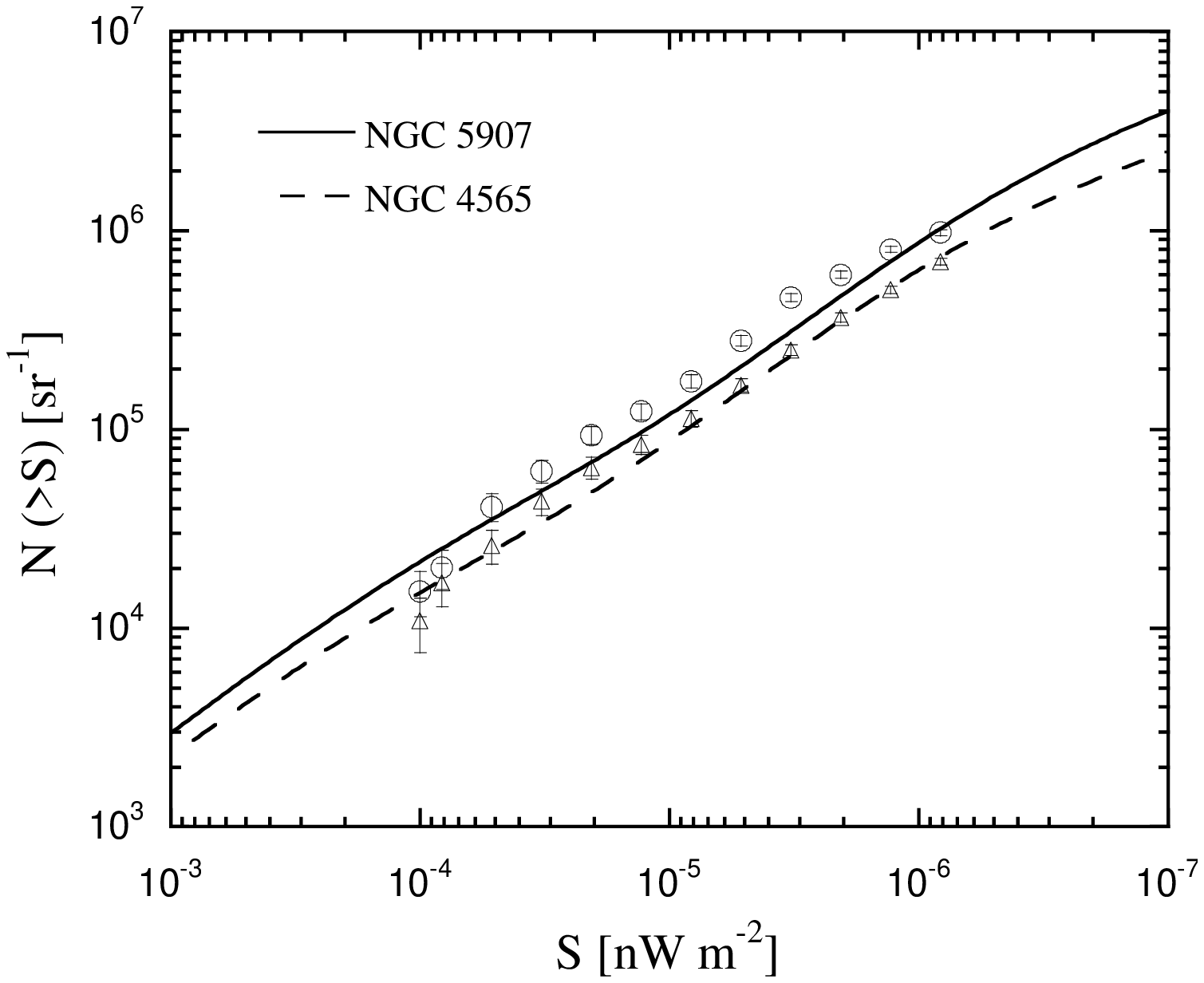}{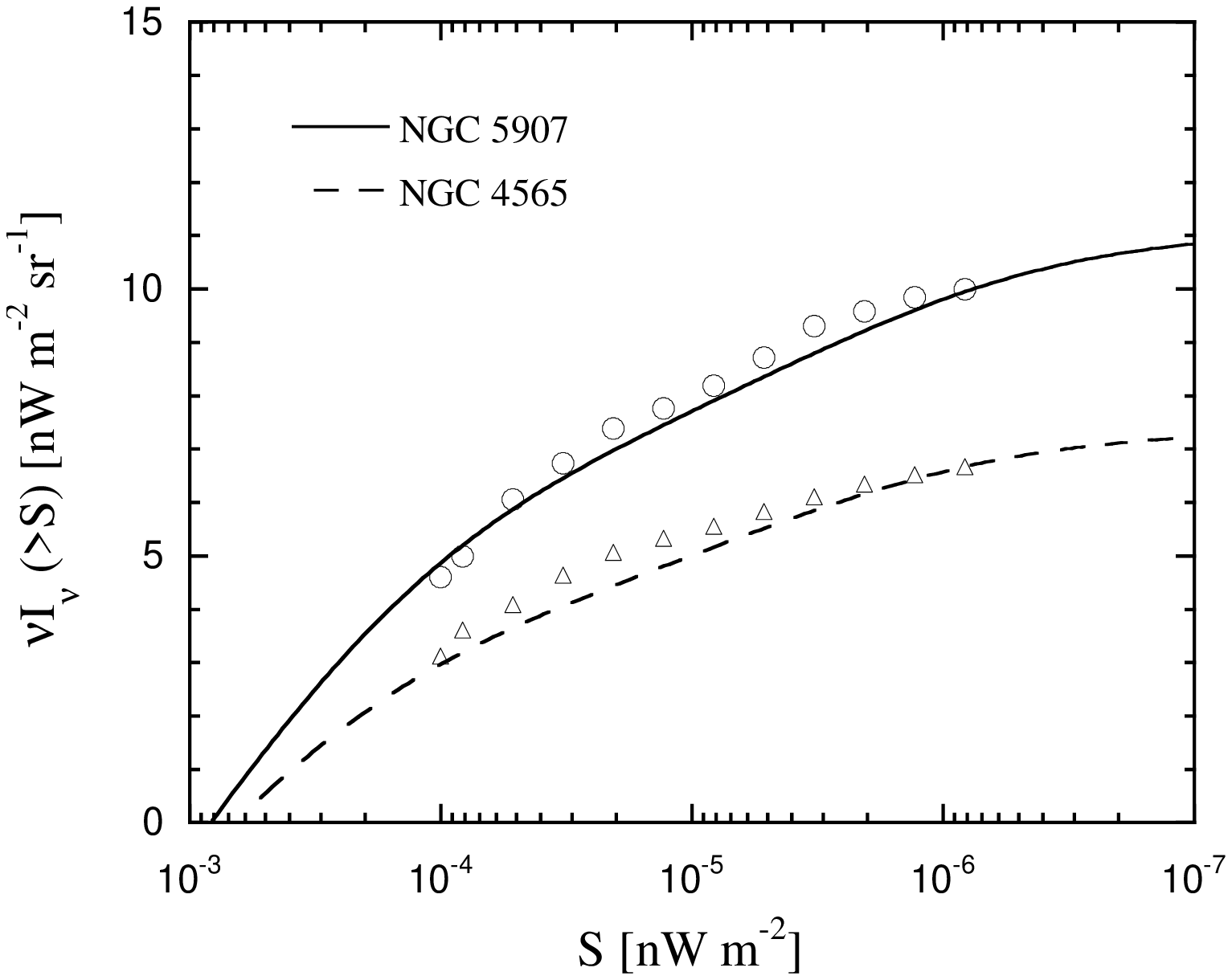}
\caption{ a) Integrated number counts are plotted for all sources detected with $> 4\sigma$ significance, shown for the NGC 4565 (\scriptsize$\triangle$\normalsize) and NGC 5907 (\tiny$\bigcirc$\normalsize) fields.  The solid curves denote predictions of the Wright and Reese (2000) stellar model.  b) Integrated surface brightness is plotted for the two fields, with data and model predictions as in Fig. 2a. \label{fig:num_count}}
\end{figure}

The source counts presented in Fig. \ref{fig:num_count} appear to be dominated by stars.  The counts may be parameterized as

\begin{equation}
N(S) = N_{0} \, {(\frac{S}{S_{0}})}^{-\alpha} ,
\label{eqn:power}
\end{equation}

\noindent
where \( N_{0} = 6.0 \times 10^{5} \) $sr^{-1}$, \( S_{0} = 1.0 \times 10^{-6} \) $nW$ $m^{-2}$, and $\alpha = 0.76$ for the NGC 4565 field.  The index of the counts, $\alpha = 0.76$, is much shallower than that of deeper $K$-band galaxy counts, \( \alpha \sim 1.7 \) \citep{sar97, ber98, sar99, jar01}.  The extended surface brightness due to detected point sources amounts to \( \nu I_{\nu} = 6.7 \) $nW$ $m^{-2}$ $sr^{-1}$ in the NGC 4565 field, and 10.0 $nW$ $m^{-2}$ $sr^{-1}$ in the NGC 5907 field.  The ratio of the surface brightness of the two fields from detected sources, \(1.0 \times 10^{-4} \ nW \ m^{-2} > \nu F_{\nu} > 6.3 \times 10^{-7} \ nW \ m^{-2} \), is 1.50, somewhat larger than that predicted by the stellar model, 1.38, but consistent with the counts being largely stars and not galaxies.  We are able to cross-identify 548 sources in the NGC 4565 field with Digitized Sky Survey images.  Of the 548 cross-identified sources, 27 appear to be galaxies based on their measured FWHM.

We can compare the source counts in the NGC 4565 and NGC 5907 fields to the counts from the model of Wright \& Reese (2000) based on the Wainscoat et al. (1992) Galactic model, hereafter referred to as the W\&R stellar model.  NITE represents a significant check on the model, probing the counts down to \( \nu F_{\nu} > 6.3 \times 10^{-7} \) $nW$ $m^{-2}$ (0.8 $mJy$) in a wide field.  As the W\&R stellar model computes differential counts in standard $L$ and $M$ bands, we note that \( N_{M}(< M) = N_{L}(< L + 0.2) \) over the range $L$ = 5 to 15.  Thus in order to estimate number counts in the NITE band, we compute \( N_{NITE}(> S) = N_{L}(< L - 0.1) \), where the flux $S$ is converted to $nW$ $m^{-2}$ in the NITE band in the same manner as the calibration stars.  The integrated source counts agree reasonably well with the model, and are shown in Fig. \ref{fig:num_count}a.  The integrated surface brightness, \( \nu I_{\nu}(> S) = \int S \, (dN / dS) \, dS \), where the flux $S$ is in $\nu F_{\nu}$ units, is compared with the measured $\nu I_{\nu}(> S)$ in Fig. \ref{fig:num_count}b.  An offset has been subtracted from the model by setting $\nu I_{\nu}(> S)$ to the measured value at \( \nu F_{\nu} = 6.3 \times  10^{-7} \) $nW$ $m^{-2}$, the source detection limit.  The offset must be subtracted to account for the surface brightness of bright, but rare, stars that increase the surface brightness of the model but were not present in the NITE field.  The W\&R stellar model predicts that the surface brightness contribution for all sources below the detection threshold amounts to \( \Delta \nu I_{\nu} = 0.7 \) $nW$ $m^{-2}$ $sr^{-1}$ for the NGC 4565 field.  Thus we estimate that our infrared images detect 90\% of the stellar surface brightness contribution.

\subsection{Surface Brightness Fluctuations} 

The combination of a good statistical understanding of sky fluctuations and steep number counts allows us to deeply probe the number counts of undetected sources.  Following Condon (1974), we may recover information on the density of sources below the detection limit, given our large statistical ensemble of 5000 pixels (Section 3.4).  We model the distribution as arising from Gaussian instrument and flat-field noise, estimated in section 3.2, and unresolved sources.  The surface brightness fluctuations due to Poissonian variations in a beam $\Omega$, where $\Omega = 9.87 \times 10^{-4}$ $sr$,

\begin{equation}
v = {\sigma_{s}}^{2} = \frac{1}{\Omega} \int_{S_{d}}^{0} {S^{'}}^{2} \,  \frac{dN}{dS^{'}} \, dS^{'}
\end{equation}

\noindent
may be used to constrain the surface brightness, $\nu I_{\nu}$, due to sources below the detection limit,

\begin{equation}
\nu I_{\nu}(>S) = \frac{1}{\Omega} \int_{ }^{S} S^{'} \,  \frac{dN}{dS^{'}} \, dS^{'}
\end{equation}

Thus for a source distribution described as a power law (eq. [\ref{eqn:power}]), fluctuations probe deeply into the background if the source counts are steep.  Sources brighter than the cutoff limit, $S_{c}$, produce a measurable change in the brightness histogram according to

\begin{equation}
\frac{{\delta}_{v}}{v} = \frac{\int_{S_{c}}^{0} S^{2} \, \frac{dN}{dS} \, dS }{\int_{S_{d}}^{0} S^{2} \,  \frac{dN}{dS} \, dS}  = {(\frac{S_{c}}{S_{d}})}^{-\alpha+2} ,
\label{eqn:cutoff}
\end{equation}

\noindent
where $S_{d}$ is the point source detection limit and $\delta_{v}$ is the change in the variance produced by all sources fainter than $S_{c}$.  Since our measurement of the fractional uncertainty in the sky variance is \( \delta_{v} /v = 10.3 \% \) ( \( v_{s} = (7.49 \pm 0.77) \) $nW^{2}$ $m^{-4}$ $sr^{-2}$, Section 3.2), the histogram is sensitive to sources much fainter than the detection limit $S_{d}$.  For example $S_{c}/S_{d} = 10^{-2}$ in the case of Euclidean counts ($\alpha = 1.5$).  If $\alpha > 2$, fluctuations will be dominated by sources where the power law relationship eventually rolls over, in principle well below $S_{d}$.

\begin{figure}
\plotone{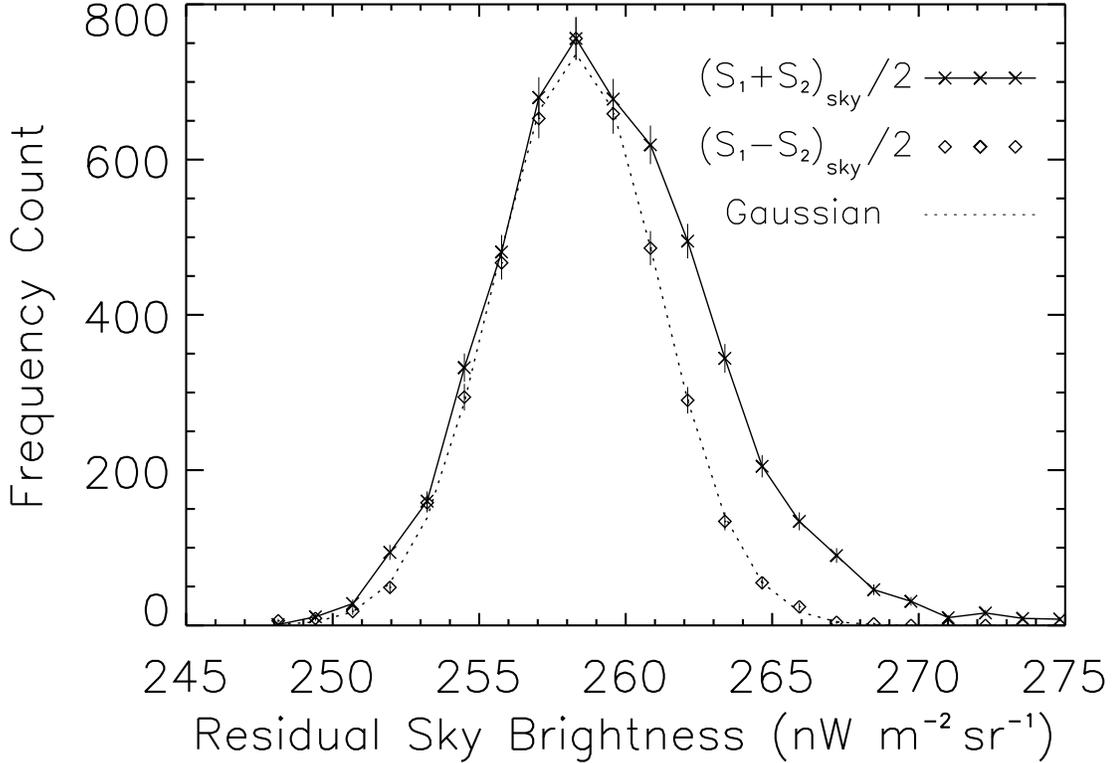}
\caption{ Brightness distributions of masked IR image about NGC 4565.  The histogram plot of the summed image ${(S_{1} + S_{2})}_{sky}/2$, $H_{S_{sky}}$, exhibits an asymmetry about the mean, and is significantly wider than that of the differenced image ${(S_{1} - S_{2})}_{sky}/2$, $H_{D_{sky}}$.  The difference histogram $H_{D_{sky}}$ is fitted to a Gaussian.  The variance in each pixel of the masked image is thus marginally dominated by sky fluctuations.  The near-Gaussian shape of $H_{S_{sky}}$ indicates a large number of sources per pixel, requiring a rapid increase in the source counts below the point source detection threshold. \label{fig:histogram}}

\end{figure}

\begin{figure}
\plotone{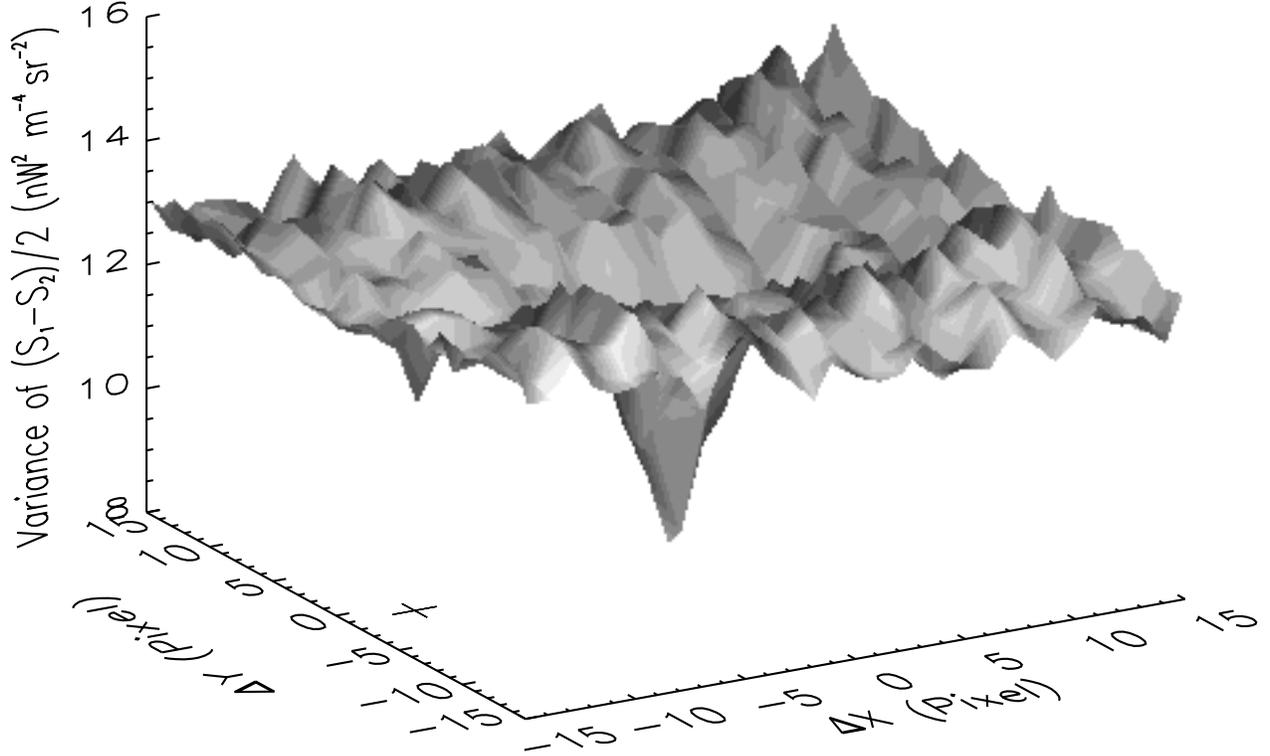}
\caption{ Variance in \( D = (S_{1} - S_{2})/2 \) is plotted as a function of positional displacement between $S_{1}$ and $S_{2}$.  The $\Delta$x - $\Delta$y coordinates correspond to the relative offsets on the sky of $S_{1}$ and $S_{2}$ .  The minimum at \( (\Delta x, \Delta y) = (0, 0) \) pixel corresponds to subtracting the images in sky coordinates, confirming the presence of a fixed pattern on the sky, or ``sky fluctuations''.  The shape of the minimum is consistent with the instrument PSF FWHM.  There is no appreciable decrease in variance at \( (\Delta x, \Delta y) = (-12, -1) \), when the two images are differenced according to array coordinates, as marked by the ``+'' in the $\Delta$x - $\Delta$y plane.  The overall fluctuation ($\sim$ 5\% in amplitude) pattern in variance evaluation is an artifact arising from omitting data obtained with noisy readout columns. \label{fig:difplot}}
\end{figure}

\begin{figure}
\plotone{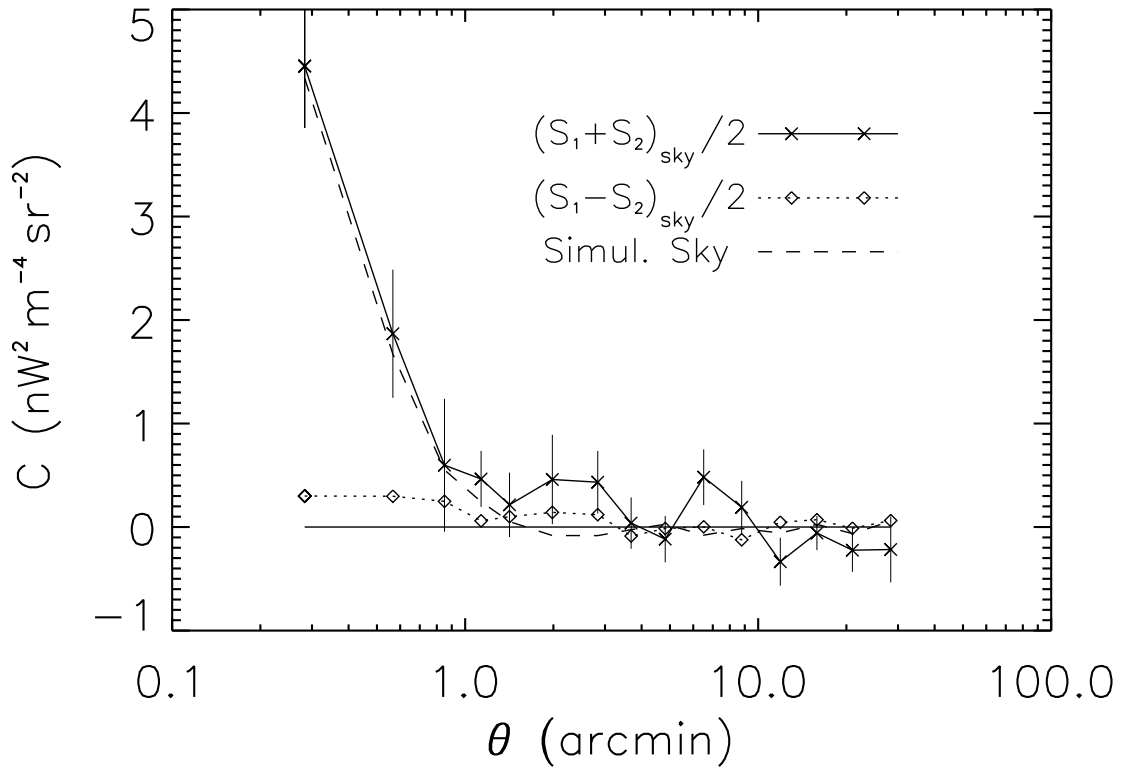}
\caption{ Auto-correlation function of the sum and difference frames.  The gradual rise in the sum frame, $C_{S_{sky}}(\theta)$ near $\theta = 0'$, is consistent with a combination of instrument noise and sky fluctuations arising from unresolved sources, spatially varying as the instrument PSF.  $C_{D_{sky}}(\theta)$ shows no gradual rise, as expected for variance arising from instrument and flat-field noise. \label{fig:autocorrelation}}
\end{figure}

\begin{figure}
\plotone{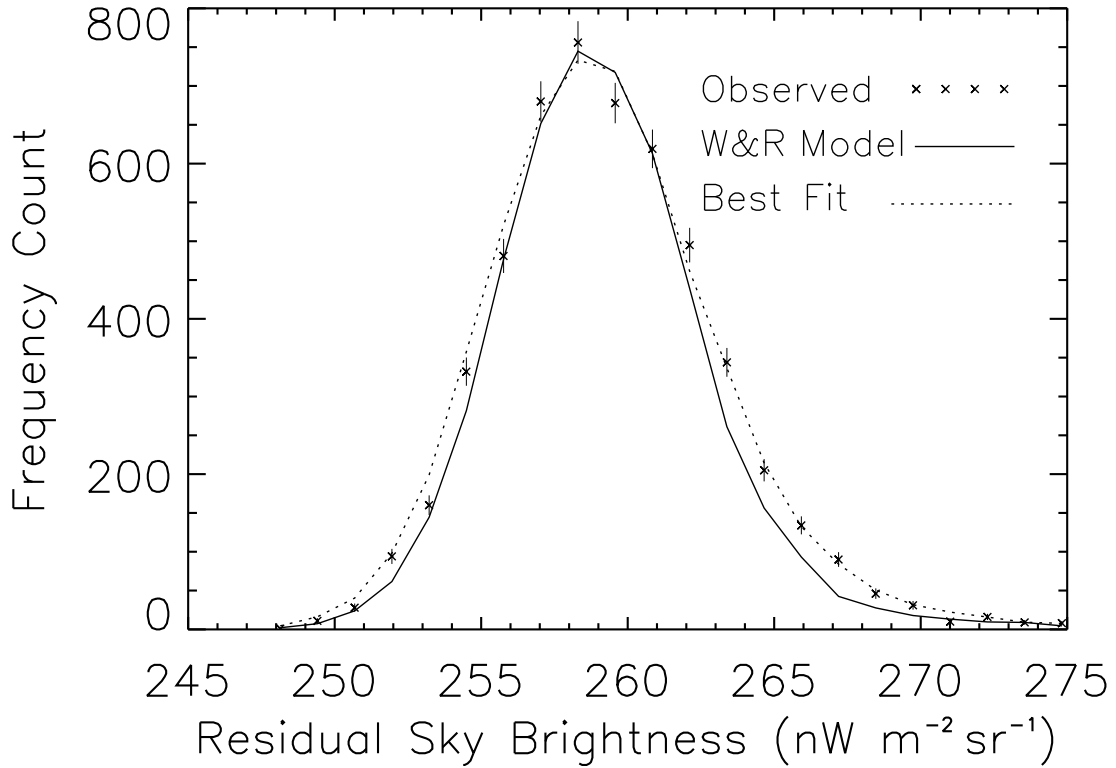}
\caption{ Data points with Poissonian error bars denote the measured histogram $H_{S_{sky}}$ of the combined sky image.  The estimated histogram for the combination of instrument noise and undetected stars taken from the W\&R stellar model results in a poor approximation to the observed fluctuations, as shown by the solid line.  Additional faint sources are required to broaden the simulated histogram.  We model the observed fluctuations with two components, stellar counts taken from the W\&R model, and a power law galaxy population as described in section 4.2.  The best-fit simulated distribution is shown by the dashed line. \label{fig:bestfit}}
\end{figure}

The width and near-Gaussian shape of the $H_{S_{sky}}$ in Fig. \ref{fig:histogram}  indicate a rapid increase in the source counts below the detection threshold.  Comparison with $K$-band galaxy counts ($\alpha \sim 1.7$) also suggests the need for higher $\alpha$.  Therefore, we fit the data to two components, stellar counts taken from the W\&R stellar model, and a power law population (eq. [\ref{eqn:power}]) with undetermined $N_{0}$ and $\alpha$.  We assume that the latter component represents galaxies.  Assuming the source counts are dominated by stars taken from the W\&R stellar model, our simulation results in a poor approximation to the observed fluctuations, as shown by the solid line in Fig. \ref{fig:bestfit}.

We carry out a P(D) analysis to estimate the density of sources below the source detection limit, and constrain the associated background produced by these sources.  We created a Monte Carlo sky to best estimate faint galaxy source count parameters $N_{0}$ and $\alpha$ (eq. [\ref{eqn:power}]).  The simulated sky contains point sources with flux \( 1.0 \times 10^{-9} \ nW \ m^{-2} \leq \nu F_{\nu} \leq 6.3 \times 10^{-7} \ nW \ m^{-2} \).  According to eq. (\ref{eqn:cutoff}), we must include all sources brighter than $S_{c}$.  Including sources fainter than $S_{c}$ has no appreciable affect on the results.  We begin our simulation with a \( 1^{\circ} \times 1^{\circ} \) field, gridded with $4.25''$ pixels. We populate the grid with randomly placed delta functions, which we generate assuming a galaxy power law source counts (eq. [\ref{eqn:power}]), and the W\&R stellar model, selected in steps of 0.1 magnitude (corresponding to a multiplicative factor of $10^{-0.04}$ $nW$ $m^{-2}$).  We then convolve the simulated image with the PSF, and re-bin to $17''$ pixels to match that of the detector array.  Finally we add Gaussian noise with dispersion $\sigma_{n}$ to each pixel to simulate instrument and flat-field noise.   

The simulated sky is then compared with the observed histograms $H_{S_{sky}}$ and $H_{D_{sky}}$ (see Fig. \ref{fig:histogram}).  We fit the histograms jointly, allowing $N_{0}$, $\alpha$, $I_{0}$, and $\sigma_{n}$ to be free parameters.  $I_{0}$ accounts for the difference in mean between the simulated and the observed skies. We must let $I_{0}$ be undetermined since our data contains a significant undetermined contribution from Zodiacal emission.  Letting $I_{0}$ vary also means that our simulation is insensitive to the faintest sources ($S < S_{c}$) applied in the Monte-Carlo simulation, since these sources add surface brightness but give no appreciable change in the histogram.  We let $\sigma_{n}$ vary to allow for uncertainty in the instrument and flat-field noise.  The quantity $\sigma_{n}$ has an experimental error (see eq. [\ref{eqn:varB}]) that affects both $H_{S_{sky}}$ and $H_{D_{sky}}$.  Fitting $H_{S_{sky}}$ and $H_{D_{sky}}$ jointly properly accounts for the uncertainty in $\sigma_{n}$, and essentially constrains $\sigma_{n}$ to values within experimental errors.

\begin{figure}
\plotone{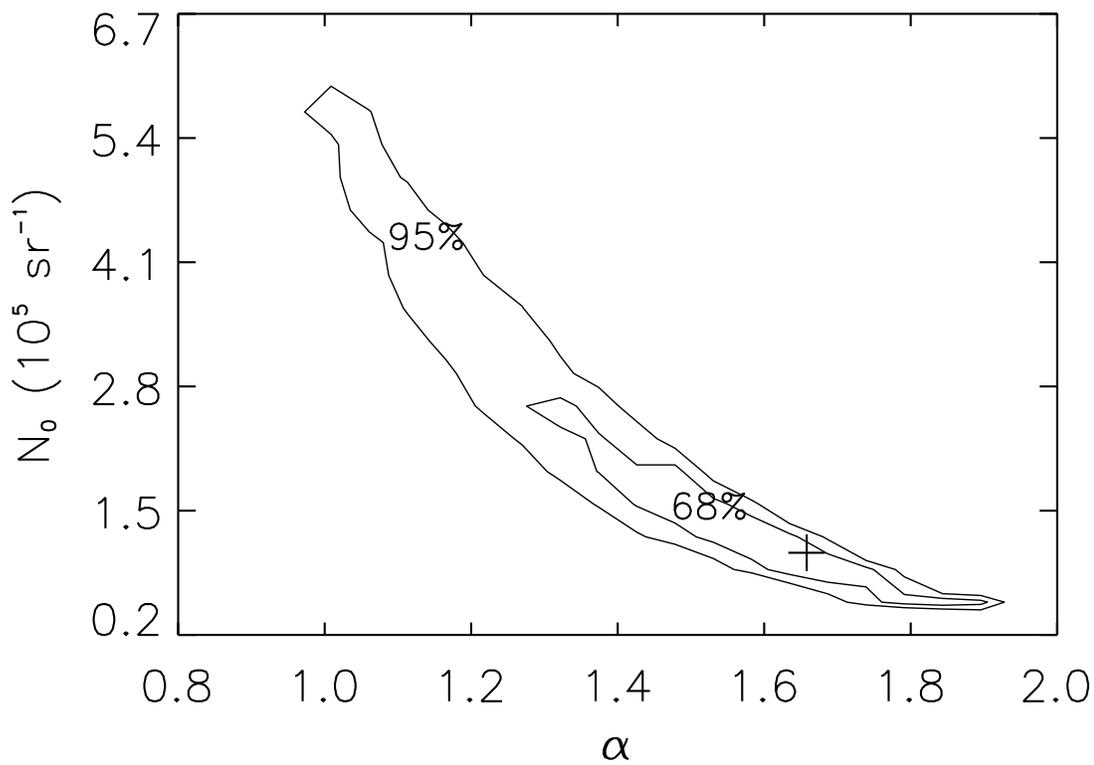}
\caption{
68\% and 95\% confindence contours for our best-fit result to the
source population producing brightness fluctuations, as described in section
4.2.  Our best-fit result, marked by ``+'', gvies $N_{0} = 0.8^{-0.3}_{+1.6} \times 10^{5}$ $sr^{-1}$, and $\alpha = 1.65^{+0.15}_{-0.35}$, with $S_{0}$ held at the numerically convenient value of 10$^{-6}$ $nW$ $m^{-2}$.
\label{fig:error_ellipse}}
\end{figure}

We determine the best fit (reduced $\chi^{2}$) to the histograms for a given set of $N_{0}$ and $\alpha$ values, marginalizing over $\sigma_{n}$ and $I_{0}$.  We semi-analytically truncate the analysis to \( \alpha \le 1.8 \), based on eq. (\ref{eqn:cutoff}), to handle the case where the histogram is sensitive to sources fainter than our minimum Monte-Carlo source flux.  We repeat simulations multiple times to accumulate statistical power and to reduce the numerical error in the simulation.  We examine the goodness-of-fit over a wide range of $N_{0}$ and $\alpha$, as shown in Fig. \ref{fig:error_ellipse}, with 50 simulations performed for each point in the $N_{0} - \alpha$ parameter space.  Our best-fit result, as shown by the dashed line in Fig. \ref{fig:bestfit}, and marked by the ``+'' in Fig. \ref{fig:error_ellipse}, gives $N_{0} = 0.8^{-0.3}_{+1.6} \times 10^{5}$ $sr^{-1}$, and $\alpha = 1.65^{+0.15}_{-0.35}$, with $S_{0}$ fixed to the numerically convenient value of $10^{-6}$ $nW$ $m^{-2}$. 

\begin{figure}
\plottwo{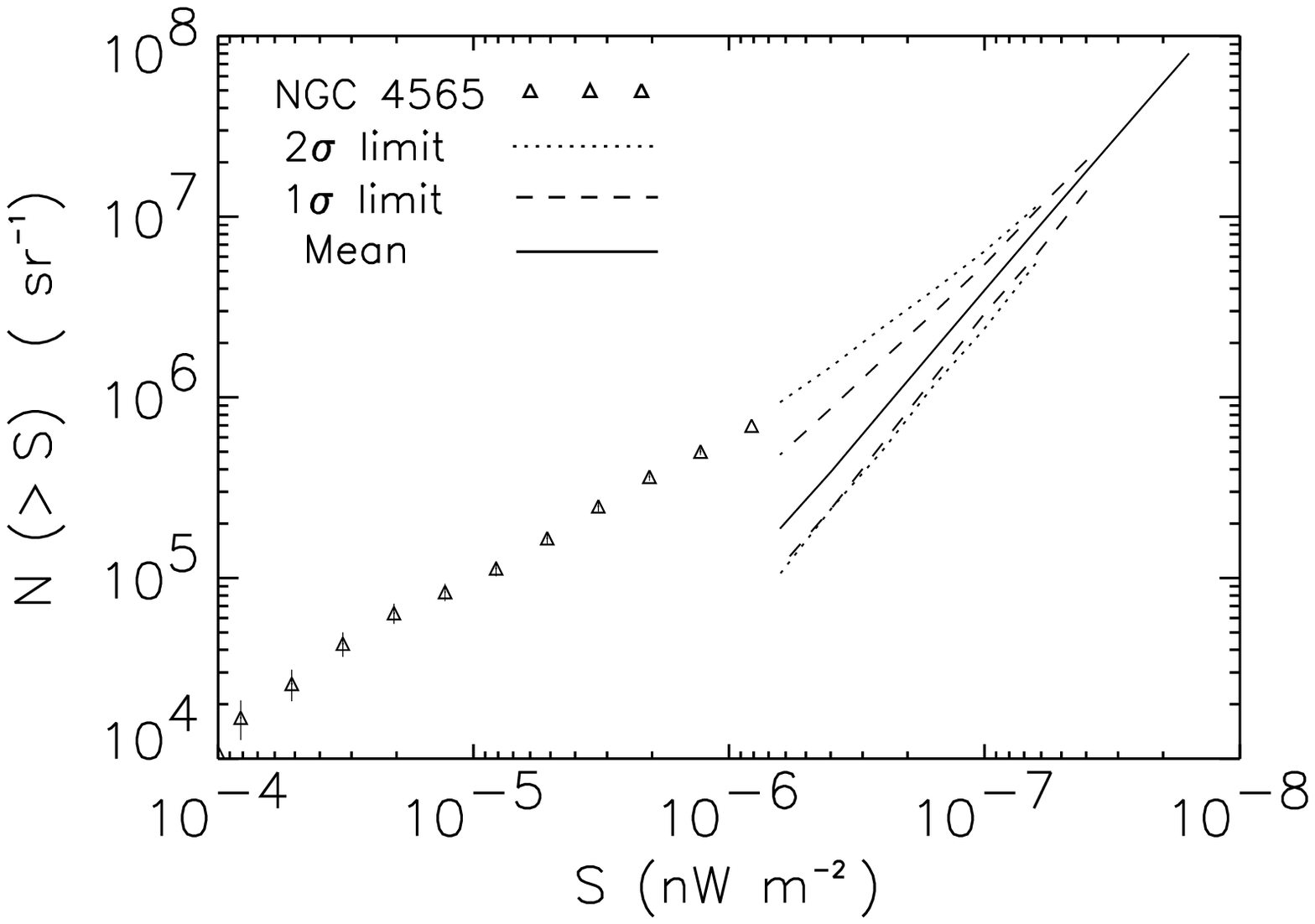}{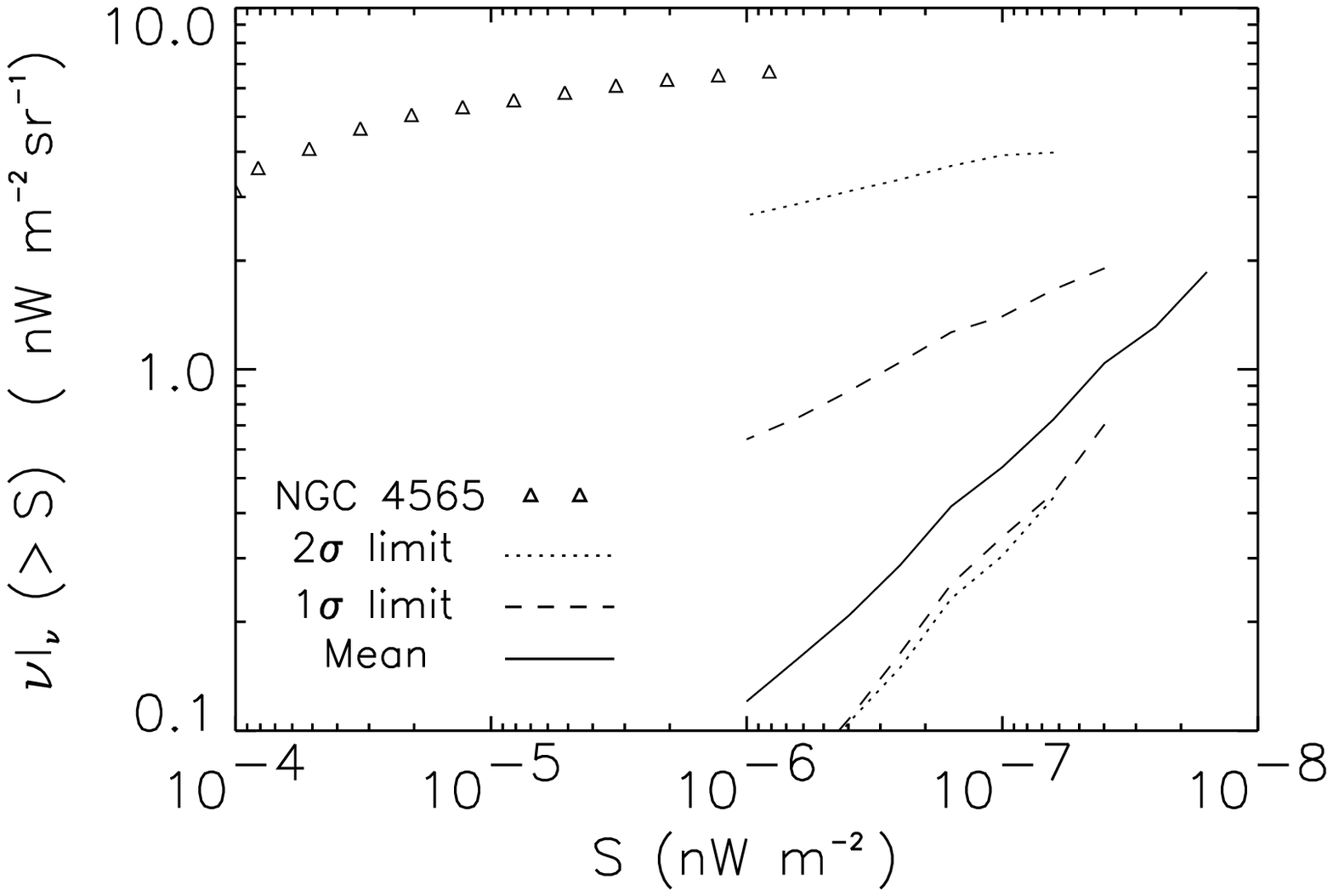}
\caption{
Estimated $N(>S)$ and $\nu I_{\nu}(>S)$ curves based on the Monte-
Carlo $P(D)$ analysis derived from the error ellipses shown in Fig. \ref{fig:error_ellipse}.  The solid curves denote the best fit result, with dashed and dotted curves indicating 68\% and 90\% confidence curves respectively.  The curves are terminated at $S_{c}$ determined from the Monte-Carlo simulation.  Since the 2$\sigma$ and 1$\sigma$ error ellipses include low values of alpha, with a higher corresponding cutoff flux, these curves are truncated at their corresponding maximum cutoff flux which is higher than the cutoff flux for the best fit curve.  Also plotted are the data points from sources extracted from the NGC 4565 field, the field used to produce the histograms for the $P(D)$ analysis.
\label{fig:error_bands}}
\end{figure}

The error ellipse shown in Fig. \ref{fig:error_ellipse} may be translated into $N(>S)$ and $\nu I_{\nu} (>S)$ vs. $S$ plots by applying eq. (\ref{eqn:power}) and (\ref{eqn:cutoff}), and propagating the errors appropriately.  We evaluate the ``true'' cutoff fluxes using a Monte-Carlo simulation, and find them to be brighter than expected from eq. (\ref{eqn:cutoff}),  presumably due to the addition of the fitted parameters $I_{0}$ and $\sigma_{n}$.  We adopt the simulated cutoff fluxes for the 1$\sigma$ and 2$\sigma$ error contours,  rather than those from eq. (\ref{eqn:cutoff}).  We truncate the error bars when any portion of the error ellipse falls below the corresponding cutoff flux.  The resulting constraints are shown in Fig. \ref{fig:error_bands}, along with the source count data from the NGC 4565 field.  The associated errors on $N(>S)$ and $\nu I_{\nu} (>S)$, also shown in Fig. \ref{fig:error_bands}, improve at fainter fluxes because the error ellipse rotates to make the projected error in $N(>S)$ and $\nu I_{\nu} (>S)$ smaller.  We thus determine the integrated source counts \( N(>S) = 1.79^{+0.26}_{-0.40} \times 10^{7} \) $sr^{-1}$, contributing \( \nu I_{\nu} (>S) = 1.04^{+0.86}_{-0.34}\) $nW$ $m^{-2}$ $sr^{-1}$ to the CIBR, evaluated at \( S = 4.0 \times 10^{-8} \) $nW$ $m^{-2}$. 

We note that this Monte-Carlo technique is limited by the statistical power of the data.  A more precise statistical determination of the histograms $H_{S_{sky}}$ and $H_{D_{sky}}$ would produce a smaller error ellipse in Fig. \ref{fig:error_ellipse}.  This would allow us to exclude the smallest values of $\alpha$ currently in the error ellipse, and consequently carry the curves in Fig. \ref{fig:error_bands} deeper since the maximum allowed cutoff flux would be lower.

The values of $N_{0}$ and $\alpha$ agree with the observed galaxy counts at $K$-band (\( N(>S_{K}) \sim 1.8 \times 10^{7} \) $sr^{-1}$ at \( S_{K} \sim 1.0 \times 10^{-7} \) $nW$ $m^{-2}$, and $\alpha \sim 1.7$; Saracco et al. 1997; Bershady et al. 1998; Saracco et al. 1999; Jarrett et al. 2001), giving an average color \( K - M_{NITE} = 1.0 \).  The $K$-band counts show a roll-off at $K \sim 17$, which is not included in our source count model.  Alternately, we can assume reddened $K$-band galaxy counts (\( K - M_{NITE} = 1.0 \)) in our simulation.  The result is consistent with the histogram data and produces a background of 2 $nW$ $m^{-2}$ $sr^{-1}$.

As a check, we fit to the histogram $H_{S_{sky}}$ and employ three point Gauss-Hermite quadrature summation \citep{zwi96, gan97} to estimate the errors arising from variation in $\sigma_{n}$.  The best-fit result is consistent with the joint analysis.

\subsection{Limits on Extragalactic Background}

DIRBE produced an absolutely calibrated all sky map at 10 different IR wavelength from 1.25 $\mu$m to 240 $\mu$m.  Unfortunately, DIRBE's $0.7^{\circ}$ beam is contaminated by Galactic foreground stars.  Since it is difficult to observe large regions of sky to sufficient depth in the thermal near-infrared, models and fits of the Galactic stellar flux have been the primary method used to remove the contamination from the DIRBE maps \citep{hau98, dwe98, wri00}.  Recently the Two Micron All Sky Survey (2MASS) provided the needed large-scale, high-resolution data to help determine the background at 1.25 $\mu$m and 2.2 $\mu$m \citep{cam01, gor00}.  Gorjian, Wright, \& Chary (2000) also determined the brightness contribution from foreground stars at \( \lambda = 3.5 \: \mu m \) in selected DIRBE fields, down to a limiting flux of \( 3.8 \times 10^{-5} \) $nW$ $m^{-2}$.  In comparison, these observations are 50 times deeper in $\nu F_{\nu}$ than Gorjian, Wright, \& Chary (2000) and cover a wider field.

Galactic dust emission appears to be negligible in our field.  Arendt et al. (1998) established an empirical relation between residual Galactic dust 100 $\mu$m emission not associated with neutral hydrogen, and emission at other wavelengths.  We use 16 DIRBE pixels on the perimeter of our field to derive a 100 $\mu$m intensity of 52.12 $nW$ $m^{-2}$ $sr^{-1}$, of which 32.32 $nW$ $m^{-2}$ $sr^{-1}$ remains after extrapolating to $N_{HI} = 0$.  The inferred contribution from dust emission is completely negligible, 0.06 $nW$ $m^{-2}$ $sr^{-1}$ at 3.5 $\mu$m and 0.09 $nW$ $m^{-2}$ $sr^{-1}$ at 4.9 $\mu$m.  Likewise the emission from the edge-on spiral galaxy in the center of our field makes a small contribution to the surface brightness.  Although technically part of the extragalactic background, we exclude the flux, \( \nu F_{\nu} = 7.4 \times 10^{-4} \) $nW$ $m^{-2}$ for NGC 4565 and \( \nu F_{\nu} = 4.2 \times 10^{-4} \) $nW$ $m^{-2}$ for NGC 5907, amounting to 0.75 and 0.43 $nW$ $m^{-2}$ $sr^{-1}$ averaged over the \( 1.8^{\circ} \times 1.8^{\circ} \) field, respectively, from the CIRB.

The image obtained by NITE provides us with a large scale, high-resolution image, and so we can use it to subtract stellar flux from a DIRBE map with zodiacal subtraction \citep{kel98}.  We align the NITE and DIRBE fields, and use the nine DIRBE pixels completely contained within the NITE image to calculate the CIRB.

In Table 1 we account for the stellar foreground in our two fields.   We remove the small flux from galaxies resolved in POSS plates from the stellar foreground.  The color of the stellar foreground produced by $5 < L < 13$ stars is \( L - M = -0.12 \) in the Wright \& Reese (2000) model (note the value for $\nu I_{\nu}(> S)$ is slightly different than for $N(> S)$).  Undetected stars, as noted in sections 4.1 and 4.2, increase the stellar surface brightness by only 10\% based on the stellar model.  Assuming this color, the total estimated stellar foreground is 11.6 $\pm$ 0.5 (4.1 $\pm$ 0.2) $nW$ $m^{-2}$ $sr^{-1}$ at 3.5 (4.9) $\mu$m for the NGC 4565 field, and 17.4 $\pm$ 0.8 (6.2 $\pm$ 0.3) $nW$ $m^{-2}$ $sr^{-1}$ at 3.5 (4.9) $\mu$m for the NGC 5907 field.

In Table 1 we estimate the residual emission, subtracting the stellar foreground from the zodi-subtracted DIRBE sky brightness in our fields.  We also account for the small contributions from detected galaxies, undetected stars, and the central edge-on galaxy.  The resulting residuals are 24.0 $\pm$ 2.7 (36.8 $\pm$ 6.0) $nW$ $m^{-2}$ $sr^{-1}$ at 3.5 (4.9) $\mu$m for the NGC 4565 field, and 14.0 $\pm$ 2.6 (22.2 $\pm$ 5.9) $nW$ $m^{-2}$ $sr^{-1}$ at 3.5 (4.9) $\mu$m for the NGC 5907 field.  The errors are dominated by the quoted zero-point errors from the zodiacal subtraction \citep{kel98}.

The large discrepancy between the NGC 4565 and NGC 5907 fields may arise from the zodiacal subtraction.  The NGC 4565 field lies at low ecliptic latitude ($\beta = 27^{\circ}$) where zodiacal residuals are more pronounced.  We confirm that a bright star, prominent in the DIRBE map and just outside of the NGC 4565 field, does not contribute significant emission to the DIRBE pixels we used to estimate the sky brightness.

The NGC 5907 residual agrees with residuals reported by several authors:   Wright \& Reese (2000), \( \nu I_{\nu} = 12 \pm 3 \) $nW$ $m^{-2}$ $sr^{-1}$ at 3.5 $\mu$m (CIRB detection), Gorjian, Wright \& Chary (2000), \( \nu I_{\nu} = 11.0 \pm 3.3 \) $nW$ $m^{-2}$ $sr^{-1}$ at 3.5 $\mu$m (tentative CIRB detection), and Dwek \& Arendt (1998), \( \nu I_{\nu} = 14 \pm 3 \) $nW$ $m^{-2}$ $sr^{-1}$ at 3.5 $\mu$m, and \( \nu I_{\nu} = 25 \pm 6 \) $nW$ $m^{-2}$ $sr^{-1}$ at 4.9 $\mu$m (tentative CIRB detection and upper limit, respectively, with updated values taken from Hauser \& Dwek 2001).  The NGC 5907 residual also agrees with the residual brightness in the DIRBE HQB regions, \( \nu I_{\nu} = 11 \pm 6 \) $nW$ $m^{-2}$ $sr^{-1}$ at 3.5 $\mu$m and \( \nu I_{\nu} = 25 \pm  8 \) $nW$ $m^{-2}$ $sr^{-1}$ at 4.9 $\mu$m, but Hauser et al. 1998 report only upper limits on the CIRB.  Although the NGC 5907 residual adds to the circumstantial evidence for a detection of the CIRB, we cannot test for isotropy in our small field.

We estimate the CIRB by accounting for the sources producing the observed surface brightness fluctuations.  Because we assume a truncated power law for the source counts, we can only constrain sources down to $S_{c}$, and thus produce a lower limit on the CIRB.  The rapidly rising source counts required to explain the histogram are very clearly due to galaxies, as expected from $K$-band survey counts.  We evaluate the detectable infrared background by integrating over sources down to the simulated 1$\sigma$ cutoff flux, \( S_{c} = 4.0 \times 10^{-8} \) $nW$ $m^{-2}$,  16.7 times below the point source detection flux, contributing $\nu I_{\nu} (>S) = 1.04^{+0.86}_{-0.34}$ $nW$ $m^{-2}$ $sr^{-1}$ to the cosmic infrared background.

\section{Conclusions}

We analyze the point source contribution to deep, wide-field 4 $\mu$m images.  Detected sources are dominated by stars, and are in general agreement with predictions from the Wright \& Reese (2000) stellar model.  We report a detection of surface brightness fluctuations detected with an ensemble of 5000 pixels.  The brightness fluctuations do not show correlated structure and appear to be dominated by galaxies.  We estimate the CIRB arising from sources that contribute to the fluctuations.  SIRTF can use sky fluctuations to study the cosmic infrared background in greater detail, probing galaxy counts to fainter limits.  Shallow SIRTF images, distributed over the sky, can remove the stellar foreground from DIRBE, allowing a test of the isotropy of the residuals.

\acknowledgments

We wish to thank the staff from the Wallops Flight Facility and White Sands Missile Range for their dedicated support during the two flights of the NITE instrument.  We also acknowledge the assistance of Eric Reese in providing us the predictions of the stellar model, and Brendan Crill for development of the telescope translation mechanism.  Steven Price, Giovanni Fazio, Judith Pipher, and William Forrest provided essential hardware and expertise for the development of the instrument.  The authors thank Bill Reach, Peter Eisenhardt, and Edward L. Wright for several useful discussions.  The COBE data sets were developed by the NASA Goddard Space Flight Center under the guidance of the COBE Science Working Group and were provided by the NSSDC.

\clearpage

\clearpage 


\begin{deluxetable}{lllllll}
\tabletypesize{\scriptsize}
\tablecaption{Estimation of the Sky Brightness and Foreground
Components }
\tablehead{
\colhead{\ } & \colhead{\ }   & \colhead{\textbf{NGC 4565}}   &
\colhead{\ } &
\colhead{\ }   & \colhead{\textbf{NGC 5907}}   &
\colhead{\ } \\     
\colhead{Component} & \colhead{3.5 $\mu$m}   & \colhead{3.5-5 $\mu$m}   &
\colhead{4.9 $\mu$m} &
\colhead{3.5 $\mu$m}   & \colhead{3.5-5 $\mu$m}   &
\colhead{4.9 $\mu$m}          

}

\startdata
DIRBE (Zodi sub.)&37.0\tablenotemark{1} \ $\pm$ (1.7)(2.1)\tablenotemark{2}&$\cdots$&41.5 $\pm$ (0.8)(5.9)&32.1 $\pm$ (1.3)(2.1)&$\cdots$&28.7 $\pm$ (0.5)(5.9)\\
ISM&0.06&$\cdots$&0.09&0.04&$\cdots$&0.06\\
Detected Stars&10.5 $\pm$ 0.5&6.7 $\pm$ 0.3&3.7 $\pm$ 0.2&15.7 $\pm$ 0.8&10.0 $\pm$ 0.5&5.6 $\pm$ 0.3\\
Undetected Stars&1.1&$\cdots$&0.4&1.7&$\cdots$&0.6\\
Detected Galaxies&0.20&0.13&0.07&$\cdots$&$\cdots$&$\cdots$\\
Galaxy\tablenotemark{3}&1.18&0.75&0.41&0.66&0.42&0.23\\

Residual&24.0 $\pm$ 2.7&$\cdots$&36.8 $\pm$ 6.0&14.0 $\pm$ 2.6&$\cdots$&22.2 $\pm$ 5.9
\
 \enddata

\tablenotetext{1}{The units for all values are in $nW \ m^{-2} \ sr^{-1}$}
\tablenotetext{2}{The first number in the parentheses is the random 1$\sigma$ error and the second one is the systematic 1$\sigma$ error due to zodiacal subtraction from \citep{kel98}}
\tablenotetext{3}{The NITE flux was converted to the $L$ and $M$ band fluxes assuming the same colors as those for the undetected stars in the Milky Way.}

\end{deluxetable}

\clearpage

\end{document}